\documentclass[12pt]{article}
\usepackage{inputenc}
\usepackage{amssymb,amsfonts,amsbsy, amsthm}

\newtheorem{thm}{Theorem}
\theoremstyle{definition}
\newtheorem{defn}[thm]{Definition}
\newtheorem{lem}[thm]{Lemma}
\newtheorem{rem}[thm]{Remark}
\newtheorem{prob}[thm]{Problem}
\newtheorem{cor}[thm]{Corollary}

\newcommand{\K}{\mathbb{K}}
\newcommand{\F}{\mathbb{F}}
\newcommand{\Fix}{\mathrm{Fix}}

\begin{document}

\title{On Ritt's decomposition Theorem in the case of finite fields}
\author{\small Jaime Gutierrez\footnote{Both authors are partially supported by Research Project MTM2004-07086 of the Spanish Ministry of Science and Technology}\\\small Dpto. de Matem\'aticas, Estad\'{\i}stica y Computaci\'on, Universidad de Cantabria\\\small E--39071 Santander, Spain\\\small jaime.gutierrez@unican.es \and \small David Sevilla\\\small Department of Computer Science, Concordia University\\\small Montreal H3G 1M8, QC, Canada\\\small dsevilla@cs.concordia.ca}

\date{}

\maketitle

\begin{abstract}
A classical theorem by Ritt states that all the complete decomposition chains
of a univariate polynomial satisfying a certain tameness condition have the
same length. In this paper we present our conclusions about the generalization
of these theorem in the case of finite coefficient fields when the tameness
condition is dropped. \emph{(Updated April 2008: see note at the beginning of
the introduction.)}
\end{abstract}

\section{Introduction}

\emph{(\textbf{Updated April 2008:} There exists a previously published
article, of which the authors were not aware, where Ritt's second theorem is
shown to be true in all characteristics, provided we have the natural and
necessary condition that no derivative vanishes identically. This goes much
beyond the condition that $p$ does not divide the degrees (which is not a
necessary condition). See additional references \cite{Zannier1993} and
\cite{Schinzel2000}. We wish to thank Pr. Zannier for this information.)}

Our starting point is the decomposition of polynomials and rational functions in one variable.
First we define the basic concepts of this topic.

\begin{defn}
Let $\K$ be any field, $x$ a transcendental over $\K$ and $\K(x)$ the field of rational functions
in the variable $x$ with coefficients in $\K$. In the set $T=\K(x)\setminus\K$ we define the
binary operation of \emph{composition} as
\[g(x)\circ h(x)=g(h(x))=g(h).\]

We have that $(T,\circ)$ is a semigroup, the element $x$ being its neutral element.

If $f=g\circ h$, we call this a \emph{decomposition} of $f$ and say that $g$ is a \emph{component
on the left} of $f$ and $h$ is a \emph{component on the right} of $f$. We call a decomposition
\emph{trivial} if any of the components is a unit with respect to decomposition.

Given two decompositions $f=g_1\circ h_1=g_2\circ h_2$ of a rational function, we call them
\emph{equivalent} if there exists a unit $u$ such that
\[h_1=u\circ h_2 \quad (\mbox{thus, } g_1=g_2\circ u^{-1}),\]
where the inverse is taken with respect to composition.

Given $f\in T$, we say that it is \emph{indecomposable} if it is not a unit and all its
decompositions are trivial.

We define a \emph{complete} decomposition of $f\in\K(x)$ to be $f=g_1\circ\cdots\circ g_r$ where
every $g_i$ is indecomposable. The notion of equivalent complete decompositions is straightforward
from the previous concepts.
\end{defn}

\begin{defn}
Given a non--constant rational function $f(x)\in\K(x)$ where $f(x)=f_N(x)/f_D(x)$ with
$f_N,f_D\in\K[x]$ and $(f_N,f_D)=1$, we define the \emph{degree} of $f$ as
\[\deg\,f=\max\{\deg\,f_N,\ \deg\,f_D\}.\]

We also define $\deg\,a=0$ when $a\in\K$.

From now on, we will use the previous notation when we refer to the numerator and denominator of a
rational function. Unless explicitly stated, we will take the numerator to be monic, even though
multiplication by constants will not be relevant.
\end{defn}

Now we can properly state the problem of decomposition of univariate rational functions, although
this will not be our main object of study.

\begin{prob}
Given a univariate rational function, decide if it is decomposable, and in the affirmative case
compute a non--trivial decomposition of the function.
\end{prob}

It is clear that the solution of this problem provides the computability of a complete
decomposition of a function if it exists.

Next, we introduce some basic results about univariate decomposition, see \cite{AGR95} for more
details.

\begin{lem}\label{prop-univ} $ $
\begin{description}
    \item[(i)] For every $f\in T$, $\deg\,f=[\K(x):\K(f)]$.
    \item[(ii)] $\deg\,(g\circ h)=\deg\,g\cdot\deg\,h$.
    \item[(iii)] $f(x)$ is a unit with respect to composition if
and only if $\deg\,f=1$, that is,
$f(x)=\displaystyle\frac{ax+b}{cx+d}$ with $a,b,c,d\in\K$ and
$ad-bc\not=0$.
    \item[(iv)] Every non--constant element of $\K(x)$ is cancelable on
the right with respect to composition. In other words, if
$f(x),h(x)\in T$ are such that $f(x)=g(h(x))$ then $g(x)$ is
uniquely determined by $f(x)$ and $h(x)$.
\end{description}
\end{lem}

We can relate decomposition and Field Theory by means of the following classical result:

\begin{thm}[L\"uroth's Theorem]\label{th-luroth}
Let $\F$ be a field such that $\K\subset\F\subset\K(x)$. Then there exists $f\in\K(x)$ such that
$\F=\K(f)$. Also, if $\F$ contains a polynomial, $f$ can be chosen to be a polynomial.
\end{thm}

\begin{proof}
See for example \cite{Lur76} for a proof in the case $\K=\mathbb{C}$, \cite{Ste10} for one in the
general case and \cite{Wae64} for an elementary one. Constructive proofs can be found in
\cite{Net95}, \cite{Sed86} and \cite{AGR95}.
\end{proof}

Now we state one of the classical Ritt's theorems (see \cite{Rit22}) about the relations among the
complete decompositions of a polynomial that satisfies a certain condition. First we have to
define that condition.

\begin{defn}
A polynomial $f\in\K[x]$ is \emph{tame} when $\mathrm{char}\;\K$ does not divide $\deg\,f$.
\end{defn}

Ritt's theorem essentially proves that all the decompositions have the same length and are related
in a rather direct way.

\begin{defn}
A \emph{bidecomposition} is a 4-tuple of polynomials $f_1,g_1,f_2,g_2$ such that $f_1\circ
g_1=f_2\circ g_2$, $(\deg\,f_1,\deg\,g_1)=1$ and $\deg\,f_1=\deg\,g_2$.
\end{defn}

\begin{thm}[Ritt's Theorem]\label{ritt1}
Let $f\in\K[x]$ be tame and let $f=g_1\circ\cdots\circ g_r=h_1\circ\cdots\circ h_s$ be two
complete decompositions of $f$. Then $r=s$, and the sequences $(\deg\,g_1,\ldots,\deg\,g_r)$,
$(\deg\,h_1,\ldots,\deg\,h_s)$ are permutations of each other. Moreover, there exists a finite
chain of complete decompositions
\[f=f_1^{(j)}\circ\cdots\circ f_r^{(j)},\ j\in\{1,\ldots,k\},\]
such that
\[f_i^{(1)}=g_i,\ f_i^{(k)}=h_i,\ i=1,\ldots,r,\]
and for each $j<k$, there exists $i_j$ such that the $j$-th and $(j+1)$-th decomposition differ
only in one of these aspects:
\begin{description}
    \item[(i)] $f_{i_j}^{(j)}\circ f_{i_j+1}^{(j)}$ and $f_{i_j}^{(j+1)}\circ
    f_{i_j+1}^{(j+1)}$ are equivalent.
    \item[(ii)] $f_{i_j}^{(j)}\circ f_{i_j+1}^{(j)}=f_{i_j}^{(j+1)}\circ
    f_{i_j+1}^{(j+1)}$ is a bidecomposition.
\end{description}
\end{thm}

\begin{proof}
See \cite{Rit22} for $\K=\mathbb{C}$, \cite{Eng41} for characteristic zero fields and \cite{FM69b} for
the general case.
\end{proof}

In this paper we will study the generalization of this result to polynomials with coefficients in
finite fields. To that end, we will also analyze the structure of intermediate fields between
$\K(f)$ and $\K(x)$. It is already known that Ritt's theorem is false when the tameness condition
is dropped, see \cite{DW74} for a counterexample.

Let $f=g(h)$. Then $f\in\K(h)$, thus $\K(f)\subset\K(h)$. Also, $\K(f)=\K(h)$ if and only if
$f=u\circ h$ for some unit $u$. This allows the following bijection among decompositions of a
function $f$ and fields between $\K(f)$ and $\K(x)$:

\begin{thm}
Let $f\in\K(x)$. In the set of decompositions of $f$ we have the equivalence relation given by the
definition of equivalence of decompositions. If we denote as $[(g,h)]$ the class of the
decomposition $f=g(h)$, the we have then the bijection:
\[\begin{array}{ccc}
\{[(g,h)]:f=g(h)\} & \longleftrightarrow & \{\F:\K(f)\subset\F\subset\K(x)\} \\
\left[(g,h)\right] & \longleftrightarrow & \F=\K(h).
\end{array}\]
\end{thm}

Thanks to the Primitive Element Theorem (see for example \cite{Lan67}), we know that for each
non--constant $f\in\K(x)$ there exist finitely many fields between $\K(f)$ and $\K(x)$. Due to the
second part of L\"uroth's Theorem, every rational decomposition of a polynomial is equivalent to a
decomposition whose components are polynomials. Therefore it suffices to care about polynomial
decomposition in this case.

In Section 2 we introduce several elementary results about univariate function fields that arise
from Galois theory. In Section 3 we present a function that is fixed by all the automorphisms of a
univariate function field over a finite field and several results related to it. In particular, we
provide an essentially new counterexample of Ritt's theorem for finite coefficient fields.

\section{The fixing group and the fixed field}

In this section we introduce several simple notions from the classical Galois theory. Let
$\Gamma(\K)=\mathrm{Aut}_\K\K(x)$ (we will write simply $\Gamma$ if there can be no confusion
about the field). The elements of $\Gamma(\K)$ can be identified with the images of $x$ under the
automorphisms, that is, with M\"obius transformations (non--constant rational functions of the
form $(ax+b)/(cx+d)\in\K(x)$), which are also the units of $\K(x)$ under composition.

\begin{defn}$ $
\begin{itemize}
    \item Let $f\in\K(x)$. We define $G(f)=\{u\in\Gamma(\K):\ f\circ u=f\}$.
    \item Let $H<\Gamma(\K)$. We define $\Fix(H)=\{f\in\K(x):\ f\circ u=f\ \forall u\in H\}$.
\end{itemize}
\end{defn}

This definitions correspond to the classical Galois correspondences (not bijective in general)
between the intermediate fields of an extension and the subgroups of its automorphism group, as
the following diagram shows:

\[\begin{array}{ccc}
  \K(x) & \longleftrightarrow & \{id\} \\
  | & & | \\
  \K(f) & \longrightarrow & G(f) \\
  | & & | \\
  Fix(H) & \longleftarrow & H \\
  | & & | \\
  \K & \longleftrightarrow & \Gamma \\
\end{array}\]

\begin{rem}
As $\K(f)=\K(f')$ if and only if $f=u\circ f'$ for some unit $u$, we have that the application
$\K(f)\mapsto G(f)$ is well--defined.
\end{rem}

We are interested in the computability of these elements, the following results solves one of the
two parts of this question.

\begin{thm}\label{gen-fixed-field}
Let $H=\{h_1,\ldots,h_m\}\subset\K(x)$ be a finite subgroup of $\Gamma$. Let $P(T)=\prod_1^m
(T-h_i)\in\K(x)[T]$. Then any non--constant coefficient of $P(T)$ generates $\Fix(H)$.
\end{thm}

\begin{proof}{Sketch of proof.}
It can be shown that $P(T)$ is the minimal polynomial of $x$ over $\Fix(H)\subset\K(x)$. Then, a
known proof of L\"uroth's theorem (see \cite{Net95}) gives the desired result. \qed
\end{proof}

The previous theorem obviously provides an algorithm to compute the fixed field for a given finite
subgroup of $\Gamma$: compute the symmetric elementary functions in $h_1,\ldots,h_m$ until a
non--constant one is found.

About the computation of the fixing group, an elementary but inefficient algorithm is given by the
resolution of the equations given by
\[f(x)-f\left(\frac{ax+b}{cx+d}\right)=0\]
in terms of $a,b,c,d$. Another algorithm (see \cite{Sev04}) combines this idea with certain
normalization of the rational function, which simplifies the equations substantially.

Next, we state several interesting properties of the fixed field and the fixing group, see
\cite{Sev04} for details.

\begin{thm}\label{H-inf-fin}
Let $H<\Gamma$.
\begin{itemize}
  \item $H$ is infinite $\Rightarrow \Fix(H)=\K$.
  \item $H$ is finite $\Rightarrow\K\varsubsetneq\Fix(H)$, $\Fix(H)\subset\K(x)$ is a normal extension, and in
particular $\Fix(H)=\K(f)$ with $\deg\,f=|H|$.
\end{itemize}
\end{thm}

\begin{thm}\label{props-fix}$ $
\begin{description}
  \item[(i)] Given a non--constant $f\in\K(x)$, $|G(f)|$ divides $\deg\,f$. Moreover, for any field $\K$ there
is always a function $f\in\K(x)$ such that $1<|G(f)|<\deg\,f$.

  \item[(ii)] $|G(f)|=\deg\,f\Rightarrow\K(f)\subseteq\K(x)$ is normal. Moreover, if the extension
$\K(f)\subseteq\K(x)$ is separable, then
\[\K(f)\subseteq\K(x)\mathrm{\ is\ normal}\Rightarrow|G(f)|=\deg\,f.\]

  \item[(iii)] Given a finite subgroup $H$ of $\Gamma$, there is a bijection between the subgroups of $H$ and
the fields between $\Fix(H)$ and $\K(x)$. Also, if $\Fix(H)=\K(f)$, there is a bijection between
the right components of $f$ (up to equivalence by units) and the subgroups of $H$.
\end{description}
\end{thm}

\begin{proof}
For the first item, we take $f=x^2\,(x-1)^2$ gives $G(f)$=\{x,1-x\}. The other ones are
straightforward. \qed
\end{proof}

\section{Finite fields}

In this section, $\K=\F_q$ where $q=p^m$ and $p=\mathrm{char}\;\F_q$, see \cite{LN97} for several
useful results. As before, we will denote $\Gamma=\Gamma(\F_q)$.

\begin{defn}
For any $\K$, $\Gamma_0=\Gamma\cap\K[x]=\{ax+b:\ a\in\K^*,\ b\in\K\}$.
\end{defn}

\begin{thm}
$\K(x)$ is Galois over $\K$ (that is, the only functions fixed by $\Gamma(\K)$ are the constants)
if and only if $\K$ is infinite.
\end{thm}

\begin{proof}
The "if" part is the first part of Theorem \ref{H-inf-fin}. The "only if" part is a consequence of
Theorem \ref{gen-fixed-field}, as $\Gamma(\K)$ is finite whenever $\K$ is finite.
\end{proof}

The interest of $\Gamma$ and $\Gamma_0$ in the case of finite fields lies in the fact that both
groups provide non--trivial fixed fields.

\begin{thm}
The fixed field for $\Gamma_0$ is generated by $(x^q-x)^{q-1}$.
\end{thm}

\begin{proof}
According to Theorem \ref{gen-fixed-field} any non--constant coefficient of
$Q(T)=\prod_{u\in\Gamma_0}(T-u)$ generates the field. But the constant term of $Q$ is precisely
$\prod_{u\in\Gamma_0}u=(x^q-x)^{q-1}$. \qed
\end{proof}

From now on, we will denote $P_q=(x^q-x)^{q-1}$.

As $\Gamma_0\subset\Gamma$, if $f$ generates the fixed field for $\Gamma$ then $f=h(P_q)$ for some
$h\in\K(x)$. Moreover, $h$ has degree $[\Gamma:\Gamma_0]=q+1$.

\begin{thm}
Let
\[h_q=(x^{q+1}+x+1)/x^q.\]
Then the rational function $f_q=h_q(P_q)$ generates $\Fix(\Gamma)$.
\end{thm}

\begin{proof}
It is easy to prove that $\Gamma_0\cup\{1/x\}$ generates $\Gamma$. As $f_q$ is a function of $P_q$
and its degree is equal to the order of the group, it suffices to show that $f_q(1/x)=f_q(x)$. A
simple computation shows that this is indeed the case: let $y=x^{q-1}$. Then $P_q(x)=y(y-1)^{q-1}$
and $P_q(1/x)=(y-1)^{q-1}/y^q$. Thus,
\[\begin{array}{cl}
 & f_q(1/x)-f_q(x)=  \\
 \\
= & \displaystyle\frac
{\displaystyle\frac{(y-1)^{q^2-1}}{y^{q^2+q}}+\frac{(y-1)^{q-1}}{y^q}+1}
{\displaystyle\frac{(y-1)^{q^2-q}}{y^{q^2}}} -
\frac {y^{q+1}(y-1)^{q^2-1}+y(y-1)^{q-1}+1} {y^q(y-1)^{q^2-q}} = \\
= & \displaystyle\frac { (y-1)^{q^2-1} + y^{q^2}(y-1)^{q-1} + y^{q^2+q} - y^{q+1}(y-1)^{q^2-1} - y(y-1)^{q-1} - 1 } { y^q(y-1)^{q^2-q} } = \\
= & \displaystyle\frac{ (y-1)^{q^2-1}(1-y^{q+1}) + (y-1)^{q-1}(y^{q^2}-y) + y^{q^2+q}-1 }{ y^q(y-1)^{q^2-q} } = \\
= & \displaystyle\frac{ (y-1)^{q^2-1}(1-y^{q+1}) + (y-1)^{q-1}((y-1)^{q^2}-(y-1)) + y^{q^2+q}-1 }{ y^q(y-1)^{q^2-q} } = \\
= & \displaystyle\frac{ (y-1)^{q^2-1}(1-y^{q+1}+(y-1)^q) - (y-1)^q + y^{q^2+q}-1 }{ y^q(y-1)^{q^2-q} } = \\
= & \displaystyle\frac{ (y-1)^{q^2-1}(1-y^{q+1}+y^q-1) - (y-1)^q + (y^{q+1}-1)^q }{ y^q(y-1)^{q^2-q} } = \\
\end{array}\]

\[\begin{array}{cl}
 = & \displaystyle\frac{ -(y-1)^{q^2}y^q - (y-1)^q + (y-1)^q(1+y+\cdots+y^q)^q }{ y^q(y-1)^{q^2-q} } = \\
= & \displaystyle\frac{ -(y-1)^{q^2}y^q + (y-1)^q(y+\cdots+y^q)^q }{ y^q(y-1)^{q^2-q} } = \\
= & \displaystyle\frac{ -(y-1)^{q^2} + (y-1)^q(1+\cdots+y^{q-1})^q }{ (y-1)^{q^2-q} } = \\
= & \displaystyle\frac{ -(y-1)^{q^2} + (y^q-1)^q }{ (y-1)^{q^2-q}
} = 0.
\end{array}\]\qed
\end{proof}

Let $f\in\F_q(x)$. Let $\mathcal{C}=\{\K: \F_q\subseteq\K\subseteq\F_q(x)\}$ and
\[\begin{array}{cccc}
\phi: & \mathcal{C} & \longrightarrow & \mathcal{C} \\
 & \F_q(f) & \rightarrow & \Fix(G(f))=\F_q(f')
\end{array}\]
which is a well--defined application. Then it is easy to check that $f'$ is a (not necessarily
proper) right--component of $f$. Also, as $G(f)\subset\Gamma$, $f'$ is a right--component of
$f_q$. Thus, $\F_q(f)\subseteq\F_q(f')$ and $\F_q(f_q)\subseteq\F_q(f')$, therefore
$\F_q(f,f_q)\subseteq\F_q(f')$.

\begin{thm}
$\F_q(f,f_q)=\F_q(f')$.
\end{thm}

\begin{proof}
Let $\F_q(f,f_q)=\F_q(m)$. Then there is a rational function $r(x,y)$ such that $r(f,f_q)=m$. For
every $u\in G(f)$, $m\circ u=r(f\circ u,f_q\circ u)=r(f,f_q)=m$. Therefore,
$m\in\Fix(G(f))=\F_q(f')\ \Rightarrow\ \F_q(m)\subseteq\F_q(f')$. The other part is
straightforward. \qed
\end{proof}

The polynomial $P_q$ has at least two different decompositions:
\[P_q=x^{q-1}\circ (x^q-x)=\left( x(x-1)^{q-1}\right) \circ x^{q-1}.\]
This gives at least two decompositions for $h_q$, both involving the component
$\displaystyle\frac{x^{q+1}+x+1}{x^q}$.

\begin{thm} $ $
\begin{description}
  \item[(i)] $\displaystyle\frac{x^{q+1}+x+1}{x^q}$ is indecomposable.
  \item[(ii)] $x^q-x$ is decomposable iff q is composite, that is, $q=p^m$ with $m\geq 2$.
  \item[(iii)] $x(x-1)^{q-1}$ is indecomposable.
\end{description}
\end{thm}

\begin{proof}
\begin{description}
  \item[(i)] We will prove that for certain units $u,v\in\F_q(x)$, the function
\[u\circ\frac{x^{q+1}+x+1}{x^q}\circ v\]
is indecomposable. In particular, let $u=x+1,v=1/(x-1)$. Then
\[u\circ\frac{x^{q+1}+x+1}{x^q}\circ v=\frac{x^{q+1}}{x-1}.\]

As the degree is multiplicative with respect to composition, and so is the difference in the
degrees of numerator and denominator (see \cite[Theorem 1.14 and Corollary 1.15]{Sev04}), there is
no possible decomposition for this function and the original function is also indecomposable.

  \item[(ii)] As $G({x^q-x})=\{x-a:\ a\in\F_q\}$ and $|G(x^q-x)|=q=\deg\ x^q-x$, by Theorem
\ref{props-fix} there is a bijection between the decompositions of $x^q-x$ and the subgroups of
its fixing group. But $G(x^q-x)$ has proper subgroups if and only if its order is composite.

  \item[(iii)] Let $q=p^m$. Let $x(x-1)^{q-1}=g(h)$ with $g=x^{p^r}+g_0,\deg\,g_0\leq p^r-1$ and
$h=x^{p^s}+h_0,\deg\,h_0\leq p^s-1$. Then
\[g\circ h=h^{p^r}+g_0\circ h=(x^{p^s}+h_0)^{p^r}+g_0\circ h=x^q+{h_0}^{p^r}+g_0\circ h\]
with $\deg\,{h_0}^{p^r}\leq q-p^r$ and $\deg\,g_0\circ h\leq q-p^s$. But
\[x(x-1)^{q-1}=x^q+x^{q-1}+\ldots+x^2+x,\]
thus either $r=0$ or $s=0$ and the decomposition is trivial. \qed
\end{description}
\end{proof}

\begin{cor}
If $q$ is not prime, $P_q$ has two complete decomposition chains of different length.
\end{cor}

As there is a bijection between the subgroups of $\Gamma_0$ and the components of $(x^q-x)^{q-1}$
on the right, we will study those subgroups in order to determine whether this polynomial has
complete decompositions of different length when $q$ is prime.

\begin{defn}
$H_0=\{x+b:\ b\in\F_q\}$.
\end{defn}

\begin{lem}
$\Gamma_0$ is the semidirect product of $H_0$ and $\{ax:\
a\in\F_q^*\}$.
\end{lem}

Let $G$ be a subgroup of $\Gamma_0$. As $H_0$ has prime order, we have two cases:

\begin{itemize}
    \item $G\cap H_0=H_0$. Then $H_0\subseteq G$. If $ax+b\in G$, then for every $b'\in\F_q$ we
have $ax+b'\in G$. In particular, $ax\in G$, and $G_0=\{a\in\F_q^*:\ ax\in G\}<\F_q^*$. But
$\F_q^*$ is cyclic of order $q-1$, thus $G_0$ is cyclic of order $m\,|\,q-1$. In this case,
$G=H_0\rtimes G_0\cong C_q\rtimes C_m$.

    \item $G\cap H_0=\{x\}$. Then for every $a\in G_0$ there exists exactly one $b\in\F_q$ such
that $ax+b\in G$, because $(ax+b)\circ(ax+b')^{-1}=x-b'+b$. As $G_0$ is cyclic, we have that $G$
is generated by some $a_0x+b_0$ where $a_0$ generates $G_0$ and $b_0\in\F_q$.
\end{itemize}

This allows to prove the following theorem.

\begin{thm}
If $q$ is prime, then all the maximal chains of subgroups of $\Gamma_0(\F_q)$ have the same
length.
\end{thm}

\begin{proof}
Let $G_0=\{x\}<G_1<\ldots<G_n=\Gamma_0(\F_q)$ be a maximal chain. Let $i\in\{1,\ldots,n\}$ be such
that $G_{i-1}\cap H_0=\{x\}$ and for all $j\geq i$, $H_0\subseteq G_j$. For each $j\geq i$ there
exists a cyclic group $C_i$ of order $m_i$ with $m_i\,|\,q-1$ such that $G_i=H_0\rtimes C_i$.
Thus, the numbers $m_i,m_{i+1},\ldots,m_n$ are a maximal chain of divisors of $q-1$ greater or
equal than $m_i$.

On the other hand, $G_{i-1}$ must be a cyclic group of order $m_i$, therefore the orders of
$G_1,\ldots,G_{i-1}$ are a maximal chain of divisors of $m_i$.

Therefore, the length of the chain $G_0,\ldots,G_n$ is equal to the number of prime factors in a
complete factorization of $q-1$ plus two. \qed
\end{proof}

\begin{cor}
The polynomial $(x^q-x)^{q-1}\in\F_q[x]$ has maximal decomposition chains of different lengths iff
$q$ is not prime.
\end{cor}

\begin{rem}
It is possible to determine all the subgroups of $\Gamma(\F_q)$ by finding all subgroups of
$GL(2,q)$. Then all chains of subgroups can be computed, finding out whether the function $f$ has
decompositions of different lengths.
\end{rem}

\section{Conclusions}

The results in the last section show some new information about the structure of decompositions of
rational functions in the finite case; it is our hope that more can be said about possible
versions of Ritt's theorems for finite fields. Also, the algorithms presented here indicate that
fast decomposition algorithms in the finite case can be achievable, by using this structure.

\end{document}